\documentclass[12pt]{iopart}

\usepackage{graphicx}

\def\erf#1{\,\mbox{erf}\,#1\,}

\def\v#1{\mbox{\boldmath $#1$}}
\def\vB{\v{B}}
\def\vr{\v{r}}
\def\vv{\v{v}}

\begin{document}

\title{Collisionless particle dynamic in an axi-symmetric diamagnetic trap}

\author{Ivan S. Chernoshtanov}

\address{11, akademika Lavrentieva prospect, Novosibirsk, 630090 Russia}
\ead{I.S.Chernoshtanov@inp.nsk.su}
\vspace{10pt}
\begin{indented}
\item[]November 2019
\end{indented}

\begin{abstract}

Particle dynamic in an axi-symmetric mirror machine with an
extremely high plasma pressure equal to pressure of vacuum magnetic
field (so-called regime of diamagnetic confinement) is investigated.
Extrusion of magnetic field from central region due to plasma
diamagnetism leads to non-conservation of the  magnetic moment and
can result in chaotic movement and fast losses of particles. The
following mechanisms can provide particle confinement for unlimited
time: absolute confinement of particles with high azimuthal velocity
and conservation of adiabatic invariant for particle moving in
smooth magnetic field. The criteria of particle confinement and
estimations of lifetime of unconfined particles are obtained and
verified in direct numerical simulation. Particle confinement time
in the diamagnetic trap in regime of gas-dynamic outflow is
discussed.

\end{abstract}

%

\section{Introduction}

Newly proposed regime of diamagnetic confinement of plasma in a
mirror machine \cite{Beklemishev16} allows us to essentially reduce
particles and energy losses from the trap and increase power density
of thermonuclear reactions. The idea is to confine plasma with
extremely high pressure equal to pressure of magnetic field of the
trap. It leads to formation of central region with sharp boundary
occupied by dense plasma with extruded magnetic field (so called
diamagnetic ``bubble''). The effective mirror ratio inside the
``bubble'' is extremely high, so in MHD approximation longitudinal
losses of plasma from inner area of the diamagnetic trap are
suppressed. In frame of the MHD approximation the plasma and energy
losses are driven by diffusion of plasma through magnetic field on
the border of the ``bubble'' \cite{Beklemishev16,Khristo19}. There
losses grow linearly with increasing ``bubble'' length and radius
and are reduced when plasma conductivity rises.

The structure of magnetic field in the diamagnetic trap is close to
Field Reversal Configuration (FRC) \cite{Steinhauer11} with zero
field reversal. So particle dynamic in the diamagnetic trap has a
lot in common with dynamic of fast ions in FRCs.

Investigation of single-particle dynamic in the diamagnetic trap is
needed for development of kinetic models of diamagnetic confinement.
Small magnitude of magnetic field results in non-conservation of
magnetic moment $mv_\perp^2/(2B)$ and can lead to chaotic behavior
and fast longitudinal losses of particles. From the other hand,
particle energy and canonical angular momentum are integrals of
motion due to stationarity and azimuthal symmetry of magnetic field.
The aim of this work is investigation of regimes of particle
confinement and lifetime of unconfined particles in the diamagnetic
trap.

The magnetic field is assumed to be fully axisymmetric later. So
influence of possible instabilities and non-accuracy of magnetic
system of the trap are neglected. Non-symmetry of magnetic system is
seems to result in slow Arnold diffusion of energy and angular
moment. In presence of dense plasma the Coulomb collisions should to
masque the slow diffusion. Mechanisms of anomalous losses due to
plasma instabilities in diamagnetic trap are addressed for future
investigations.

The article is organized as follow. Hamilton function of particle in
the trap is discussed in the second section. Two mechanisms can
provide particle confinement in unlimited time if collision
scattering is absent, namely so-called absolute confinement
\cite{Ming1985} and conservation of radial adiabatic invariant.
There mechanisms are discussed in the third section. The estimations
for lifetime of unconfined particles and plasma confinement time in
gas-dynamic regume are found in the fourth and fifth sections.
Simple numerical example is presented in the sixth section. The
results are discussed in the Conclusion.

\section{Hamiltonian}

We consider particle dynamic in the axisymmertic magnetic field
without spirality. This field can be described by only one function,
namely magnetic flux $\Psi(r,z)=rA_\theta(r,z)$, here $A_\theta$,
$r$ and $z$ are azimuthal component of vector potential, radial and
longitudinal coordinate. The Hamilton function can be written in the
following form
\begin{equation}
H(p_r,p_\theta,p_z;r,\theta,z)=\frac{p_r^2}{2m}+\frac{(p_\theta-e\Psi(r,z)/c)^2}{2mr^2}+\frac{p_z^2}{2m}+e\varphi(r,z),
\label{ham00}
\end{equation}
here $\varphi(r,z)$ is electrostatic ambipolar potential. The
particle energy and canonical angular momentum $p_\theta$ are
integrals of motion due to stationarity and azimuthal symmetry. If
structure of magnetic field is ``bubble''-like that magnetic field
(and magnetic flux) is small inside region in central part of the
trap. Outside this region magnetic field is approximately vacuum
(examples of structure of magnetic field are shown in
\cite{Beklemishev16,Khristo19}, see also figures \ref{numN} and
\ref{numB}). We assume that the ambipolar potential is approximately
constant in region with very small magnetic field and that
longitudinal gradient scale of the potential on ``bubble'' boundary
coincides with longitudinal gradient scale of magnetic field.

It is convenient in numerical simulations to express all dimension
quantities via particle mass, vacuum magnetic field in the center of
the trap $B_0$ and cyclotron frequency calculated by vacuum magnetic
field $\Omega\equiv eB_0/(mc)$. Only quantities with dimension of
length remain after this expressing. So integrals of motion with
dimension of length will be used later together with energy and
angular momentum, namely Larmor radius calculated by full energy and
vacuum magnetic field $\rho\equiv(2\varepsilon/(m\Omega^2))^{1/2}$
and minimal possible distance between particle and axis in case of
zero magnetic field $r_{\min}=p_\theta/\sqrt{2m\varepsilon}$.

\begin{figure}[!h]
\begin{center}
\includegraphics[width=0.3\textwidth]{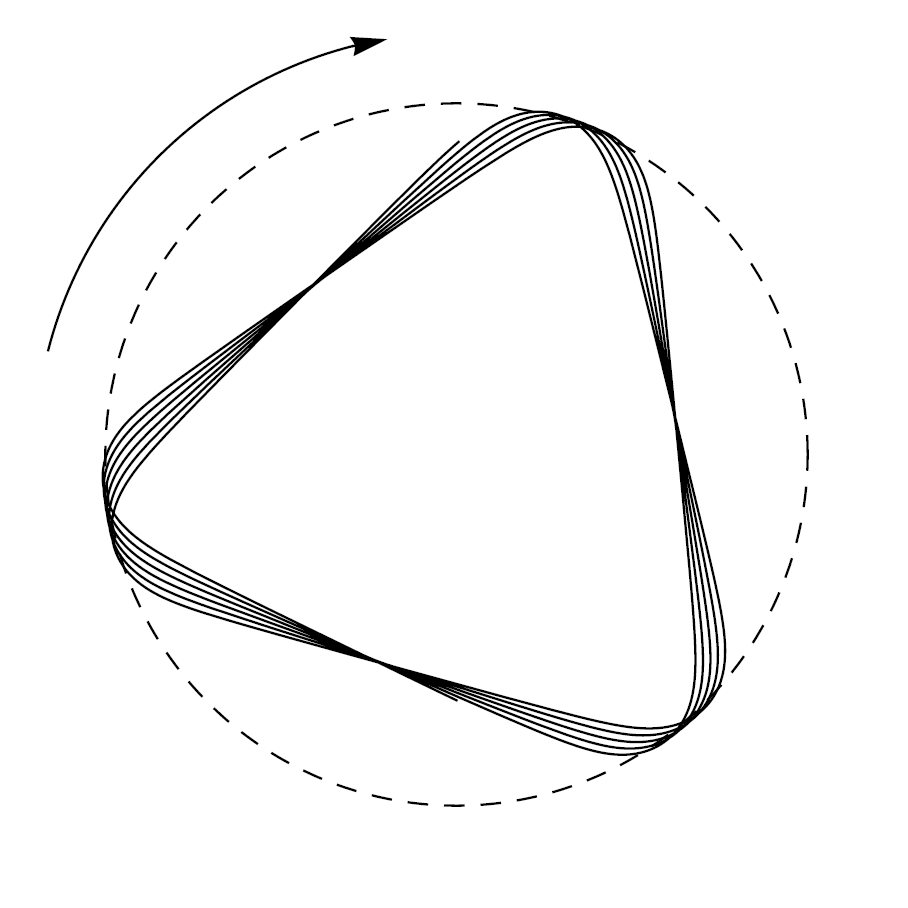}
\includegraphics[width=0.3\textwidth]{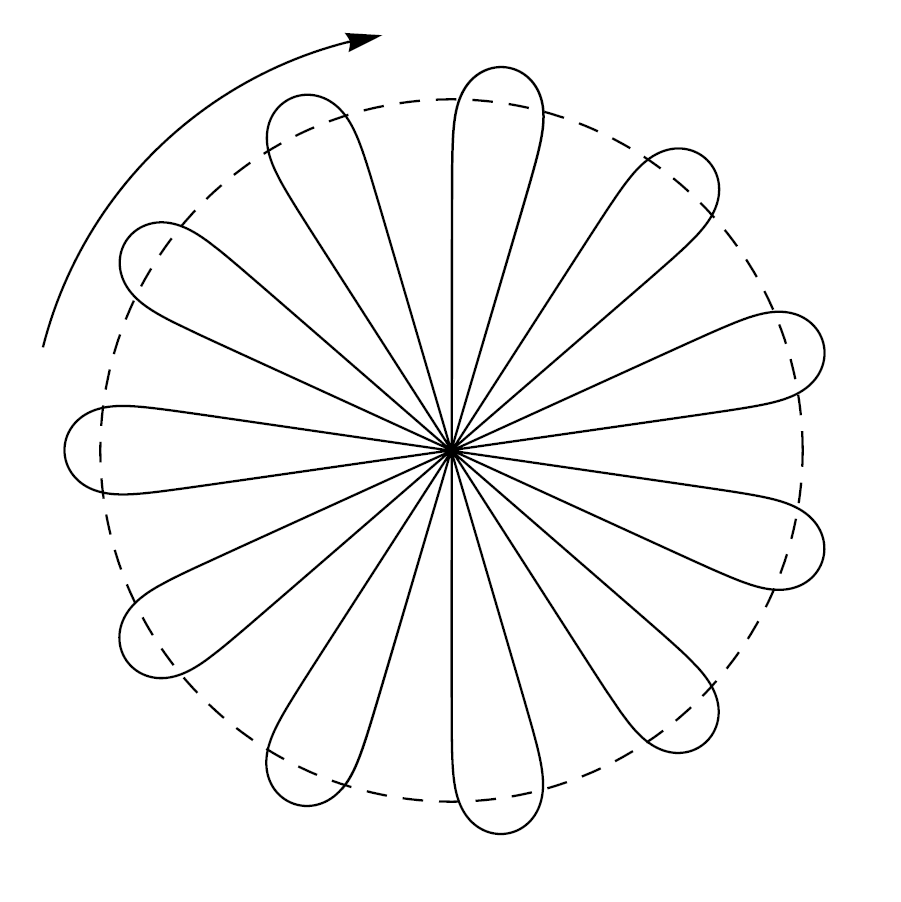}
\includegraphics[width=0.3\textwidth]{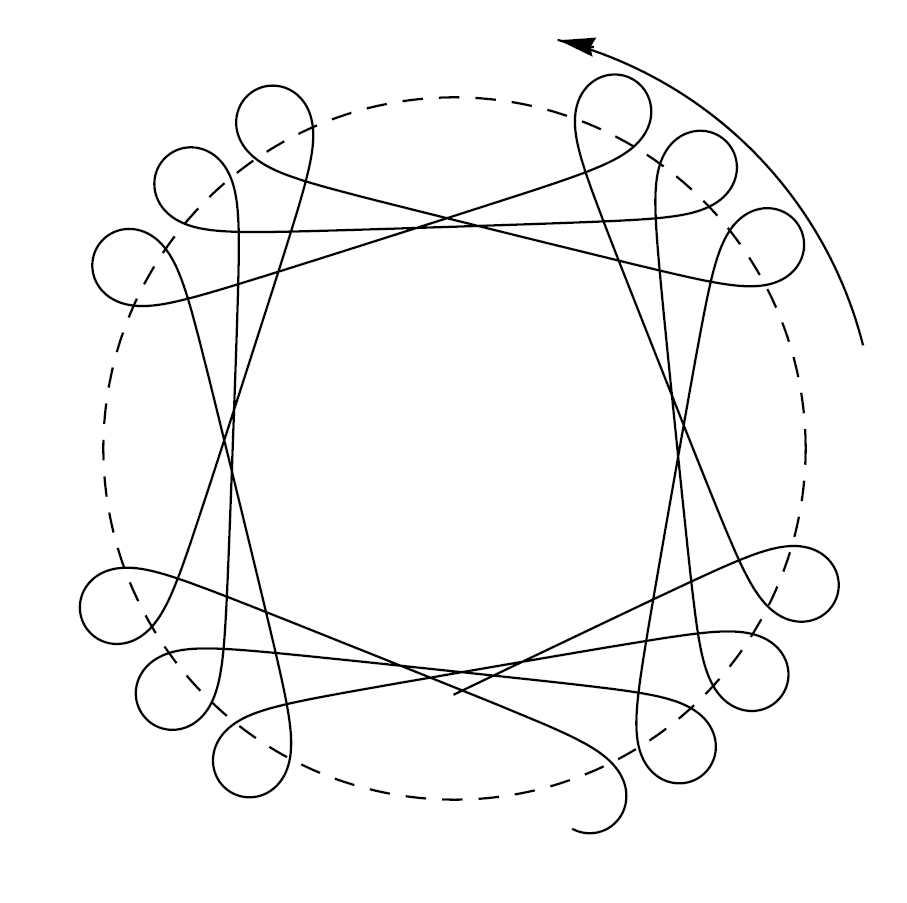}
\end{center}
\caption{\label{TraMotion} An example of trajectories of particles
with $\Omega p_\theta<0$ (left), $\Omega p_\theta=0$ (center) and
$\Omega p_\theta>0$ (right). Dashed circle bounds region $r<a$ where
magnetic field is less than $B_0/2$. Arrows indicate direction of
mean azimuthal velocity.}
\end{figure}

An example of trajectories in the transversal cross-section are
shown in figure \ref{TraMotion}. Radial dependence of the magnetic
field is chosen in the form $B(r)=B_0\{1+\tanh((r-a)/\Delta r)\}/2$,
here $a/\Delta r=10$. The trajectories can be divided into three
classes in dependence on sign of particle angular momentum. Orbits
of the particles with $\Omega p_\theta\leq0$ and $\Omega p_\theta>0$
correspond to betatron and drift orbits in FRC \cite{Rostoker02}. As
in the FRC, confinement of particles essentially depends on sign and
magnitude of angular momentum (see below).

\section{Regimes of particle confinement}

Non-conservation of magnetic momentum due to smallness of magnitude
of magnetic field inside the diamagnetic ``bubble'' results in
changing of regimes of particle confinement and essentially
modification of concept of loss cone. Two mechanisms can provide
particle confinement in the diamagnetic trap for unlimited time (in
absence of collision scattering and non-axisymmetry): well-known
absolute confinement \cite{Ming1985} and conservation of radial
adiabatic invariant.

\subsection{Absolute confinement}

The mechanism of the absolute confinement is follow. The Hamiltonian
(\ref{ham00}) described two-dimensional motion in effective
potential $(p_\theta-e\Psi(r,z)/c)^2/(2mr^2)+e\varphi(r,z)$. This
potential is potential well for particles with $\Omega p_\theta<0$.
Such particles confines in the trap if their energy is small enough.

To find condition of the absolute confinement one can note that in
region of mirrors the magnetic field if quasi-uniform,
$\Psi(r)\approx R_vB_0r^2/2$, here $R_v$ is vacuum mirror ratio of
the trap. Let's found minimal possible energy $\varepsilon_{\min}$
of particle moving with angular momentum $p_\theta$ in the mirror
region. Minimal value of effective potential is reached at a point
with radial coordinate $r_\beta$ which satisfies equation
$(-p_\theta/r_\beta^2-R_v\Omega/2)(p_\theta/r_\beta-R_v\Omega
r_\beta/2)/m+e\varphi_m'(r_\beta)=0$, here $\varphi_m(r)$ is radial
distribution of ambipolar potential in mirror. Minimal energy is
$\varepsilon_{\min}=(p_\theta/r_\beta-m\Omega
r_\beta/2)^2/(2m)+e\varphi_m(r_\beta)$. Particle with angular
momentum $p_\theta$ cannot penetrate in region of the mirror if
their energy less than $\varepsilon_{\min}$.

In simplest case of zero electrostatic potential
$r_\beta=(-2p_\theta/(m\Omega))^{1/2}$ and criterion of absolute
confinement can be written in the form \cite{Ming1985}
\begin{equation}
\varepsilon>-R_v\Omega p_\theta.\label{abs00}
\end{equation}
This criterion can be re-written in another form:
$r_{\min}>\rho/(2R_v)$ and $\Omega p_\theta<0$. Particle is confined
absolutely if it rotates quickly in the direction which coincides
with direction of cyclotron rotation.

Preferential confinement of ions with negative angular momentum can
results in spontaneous rotation of plasma in the trap which is
similar to particle-loss spin-up in FRCs \cite{Steinhauer11}.

\subsection{Adiabatic confinement}

Radial adiabatic invariant
\begin{equation}
I_r=\frac{1}{2\pi}\oint
p_rdr=\frac{1}{2\pi}\oint\sqrt{2m\varepsilon-2me\varphi-p_z^2-(p_\theta-e\Psi/c)^2/r^2}
\label{Ir00}
\end{equation}
conserves if magnetic field changes in longitudinal direction
smoothly and particle longitudinal velocity is not too large. In
this case frequency of radial oscillations
\begin{equation}
\Omega_r=2\pi/\left(\oint\frac{mdr}{\sqrt{2m\varepsilon-2me\varphi-p_z^2-(p_\theta-e\Psi/c)^2/r^2}}\right)^{-1}
\label{Ir01}
\end{equation}
can be much greater than inverse time of varying of magnetic field
during particle longitudinal motion. It should be noted that regular
motion of ions in oblate FRCs due to conservation of adiabatic
invariant is observed also in numerical simulations of FRCs (see,
for example, \cite{Belova06}).

If vacuum magnetic field has one local minimum (corrugation of field
is absent) than character time of varying of magnetic field during
particle longitudinal motion is ratio of distance between mirrors to
longitudinal velocity $L/v_\|$ so criterion of adiabaticity is
$\Omega_r>v_\|/L$. Let us consider case of large radius of
diamagnetic ``bubble'' $a\gg\rho$. In this case it is convenient to
introduce radial coordinate of magnetic field line on the ``bubble''
boundary $r_b(z)$. To choose this field line the condition
$B_z(r_b(0),0)=B_0/2$ is used. If $r_b\gg\rho$ and electrostatic
potential is zero one can calculate $\Omega_r=\pi
v_\perp^2/(v_\perp^2-\Omega^2\rho^2r_{\min}^2/r_b^2)^{1/2}$ (here
$v_\perp=(2\varepsilon/m-v_\|^2)^{1/2}$) and write criterion of
adiabaticity in the form
\begin{equation}
\frac{v_\perp}{v_\|}>\left(\frac{dr_b}{dz}\right)_{\max}\sqrt{1-\frac{\Omega^2\rho^2}{v_\perp^2}\frac{r_{\min}^2}{r_b^2}}.
\label{Ir02}
\end{equation}
Expression $(dr_b/dz)_{\max}$ denotes maximal value of derivative of
function $r_b(z)$. We assume that radial distribution of ambipolar
potential is approximately constant inside the ``bubble'' and that
longitudinal gradient scale of ambipolar potential is of the order
of $r_b^{-1}(dr_b/dz)$. In this case taking the electrostatic
potential into account does not changes criterion (\ref{Ir02})
essentially.

If motion of particle is regular than particle is confined in the
trap if real solutions $P$ of equation
$I_r(\varepsilon,p_\theta,p_z=P,z=0)=I_r(\varepsilon,p_\theta,p_z=0,z=z_m)$
are absent (here $\pm z_m$ are coordinates of mirrors). Magnetic
flux in the mirror is approximately equal to flux of vacuum magnetic
field $\Psi\approx R_vB_0r^2/2$. So adiabatic invariant (\ref{Ir00})
in the mirror is equal to
$(\varepsilon-p_z^2/(2m)+e\varphi_m(\sqrt{2|p_\theta/(m\Omega)|}))/(R_v|\Omega|)-|p_\theta|H(-\Omega
p_\theta)$, here $H(x)$ is Heaviside step function and
$\varphi_m(r)$ is radial distribution of ambipolar potential in
mirror (we assume that Larmor radius of particle in mirror is small
in comparison with radial scale length for potential). Criterion of
confinement of regularly moving particle can be written in the form
\begin{equation}
I_r(\varepsilon,p_\theta,p_z,z=0)+|p_\theta|H(-\Omega
p_\theta)>\frac{\varepsilon+e\varphi_m(\sqrt{2|p_\theta/(m\Omega)|})}{R_v|\Omega|}.
\label{Ir03}
\end{equation}
Criterion (\ref{Ir03}) can be written in analytical form if radius
of the ``bubble'' is large, $r_b(z)\gg\rho$, and ambipolar potential
is approximately constant inside the ``bubble''. In this case the
invariant (\ref{Ir00}) is equal approximately to the radial
adiabatic invariant for particle moving inside long cylinder surface
with radius $r_b(z)$:
\begin{equation*}
I_r(\varepsilon,p_\theta,p_z,z)\approx2|p_\theta|\left(\eta+\arctan\eta\right),\quad\eta=\sqrt{r_b^2(z)(2m\varepsilon-p_z^2)/p_\theta^2-1}.
\label{Ir04}
\end{equation*}

\subsection{Criterion of adiabaticity in corrugated field}

Discrete structure of magnetic system leads to corrugation of vacuum
magnetic field which results in corrugation of margin of diamagnetic
bubble. So time of essential changing of magnetic field during
longitudinal motion essentially decreases and can become comparable
with period of radial oscillations (such effect for particles moving
in vacuum magnetic field is described in \cite{Chirikov60}).
Resonant interaction between radial oscillations and longitudinal
motion can destroy adiabatic invariant (\ref{Ir00}) even if
criterion of adiabaticity (\ref{Ir02}) is satisfied. In this section
we will estimate magnitude of corrugation of vacuum magnetic field
at which corrugation not influence on movement of particle.

Let's look a longitudinally-uniform diamagnetic trap with weak
corrugation of vacuum magnetic field. Flux of vacuum magnetic field
is $\Psi_v=B_0\left\{r^2/2+(\delta b/k)rI_1(kr)\cos(kz)\right\}$,
vacuum magnetic field at $r=0$ is $B_0\{1+\delta b\sin(kz)\}$. Flux
of field in the trap is the sum $\Psi=\Psi_0(r)+\Psi_1(r)\cos(kz)$,
unperturbed part  satisfies integral equation \cite{Rostoker02}
\begin{equation}
\Psi_0(r)=B_0\frac{r^2}{2}-\frac{4\pi}{c}\int
g(r;r_0)j_\varphi(\Psi_0(r_0),r_0)dr_0, \label{corr00}
\end{equation}
here $j_\varphi(\Psi,r)$ is azimuthal component of plasma current
(which depends on distribution function of ions and electrons),
\begin{equation*}
g(r,r_0)=\frac{r^2}{2}\left(\frac{r_0^2}{r_w^2}-1\right)H(r_0-r)+\frac{r_0^2}{2}\left(\frac{r^2}{r_w^2}-1\right)H(r-r_0)
\label{corr01}
\end{equation*}
is Green function (magnetic flux generated by electric current
flowing on a cylindrical surface with radius $r_0$), $r_w$ is radius
of conducting shell surrounding plasma, $H(x)$ is Heaviside step
function.

Perturbed part $\Psi_1(r)$ satisfies following linear integral
equation:
\begin{equation*}
\Psi_1(r)=\delta bB_0\frac{r}{k}I_1(kr)-\frac{4\pi}{c}\int
g_1(r;r_0)\frac{\partial
j_\varphi(\Psi_0,r_0)}{\partial\Psi_0}\Psi_1(r_0)dr_0,
\label{corr02}
\end{equation*}
with the Green function (which is solution of the equation
$r\partial_r(r^{-1}\partial_rg_1)-k^2g_1=r\delta(r-r_0)$)
\begin{eqnarray*}
g_1(r;r_0)=rr_0\frac{I_1(kr)}{I_1(kr_w)}(I_1(kr_0)K_1(kr_w)-I_1(kr_w)K_1(kr_0))H(r_0-r)+\nonumber\\
+rr_0I_1(kr_0)K_1(kr_w)\left(\frac{I_1(kr)}{I_1(kr_w)}-\frac{K_1(kr)}{K_1(kr_w)}\right)H(r-r_0).
\label{corr03}
\end{eqnarray*}

The hamiltonian of particle moving in longitudinally-uniform
diamagnetic trap with weak corrugation is
\begin{equation*}
H(p_r,p_\theta,p_z;r,\theta,z)=\frac{p_r^2}{2m}+\frac{(p_\theta-e\Psi_0(r)/c)^2}{2mr^2}+\frac{p_z^2}{2m}-\frac{e\Psi_1(r)}{mc}\frac{(p_\theta-e\Psi_0(r)/c)}{r^2}\cos(kz)
\label{corr04}
\end{equation*}

One can make canonical transformation to the action-angle variables
of non-perturbed hamiltonian:
\begin{equation*}
H(I_r,p_\theta,p_z;\phi,\theta,z)=H_\perp(I_r,p_\theta)+\frac{p_z^2}{2m}-\frac{e\Psi_1(r(I_r,\phi,p_\theta))}{mc}\frac{(p_\theta-e\Psi_0(r(I_r,\phi,p_\theta))/c)}{r(I_r,\phi,p_\theta)^2}\cos(kz),
\label{corr05}
\end{equation*}
here $I_r(\varepsilon_\perp,p_\theta)=(2\pi)^{-1}\oint dr/(mV_r(r))$
is radial adiabatic invariant, $\phi=\{\int dr/V_r(r)\}\{\oint
dr/V_r(r)\}^{-1}$,
$V_r(r)=m^{-1}\sqrt{2m\varepsilon_\perp-(p_\theta-e\Psi_0(r)/c)^2/r^2}$
is radial velocity, $H_\perp(I_r,p_\theta)$ is solution of equation
$I_r(H_\perp,p_\theta)=I_r$. Radial coordinate $r$ depends
periodically on new variable $\phi$ (because $r$ depends
periodically on time) so perturbation of hamiltonian can be expanded
in a Fourier series:
\begin{equation}
H(I_r,p_\theta,p_z;\phi,\theta,z)=H_\perp(I_r,p_\theta)+\frac{p_z^2}{2m}-\cos(kz)\sum_n\delta
H_n(I_r,p_\theta)e^{in\phi}. \label{corr06}
\end{equation}
Condition of resonance between radial oscillations and longitudinal
motion is $n\dot\phi=n\Omega_r(I_r,p_\theta)=kp_z/m$.

Concrete form of the distribution functions of ions and electrons
are needed to calculate magnetic flux and coefficients $\delta H_n$
in (\ref{corr06}) and to follow investigation of adiabacity.
Analytical treatment can be extended in the following cases: if
angular momentum of particle is negative and small enough and if
radius of the diamagnetic bubble exceeds essentially particle Larmor
radius $\rho$.

\subsubsection{Particles with $\varepsilon\ll-\Omega p_\theta$.}

Now we consider particles with negative and very small angular
momentum. Ions with such momentum can arises in the trap due to
off-axis NBI. This ions move along betatron orbits with small
harmonic oscillations in radial direction. The unperturbed part of
hamiltonian can be written as \cite{IonRing91}
\begin{equation*}
H_\perp(p_r,p_\theta;r,\theta)=\frac{p_r^2}{2m}+\frac{(p_\theta-e\Psi_0(r_\beta)/c)^2}{2mr_\beta^2}+\frac{m\Omega_\beta^2(r_\beta)(r-r_\beta)^2}{2},
\label{corP00}
\end{equation*}
here $r_\beta$ is solution of equation
$p_\theta-e\Psi_0(r_\beta)/c+m\Omega_c(r_\beta)r_\beta^2=0$ (mean
radius of betatron orbit),
$\Omega_\beta(r)=\{\Omega_c(r)\partial_r(r\Omega_c(r))\}^{1/2}$ is
betatron frequency, $\Omega_c(r)=r^{-1}\partial_r\Psi_0(r)$ is local
cyclotron frequency. Amplitude of betatron oscillations is assumed
to be small, $|r-r_\beta|\ll r_\beta$.

Transition to the angle-momentum variables describes by canonical
transformation
\begin{equation*}
p_r=\sqrt{2m\Omega_\beta(r_\beta)I_r}\cos\phi,\quad
r=r_\beta+\sqrt{\frac{2I_r}{m\Omega_\beta(r_\beta)}}\sin\phi.
\label{corP01}
\end{equation*}
After transition to the angle-momentum variables the Hamiltonian
transforms to
\begin{equation}
H(I_r,p_\theta,p_z;\phi,\theta,z)=\Omega_\beta(r_\beta)I_r+\frac{p_z^2}{2m}+\frac{(p_\theta-e\Psi_0(r_\beta)/c)^2}{2mr_\beta^2}+\frac{e\Psi_1(r_\beta)}{c}\Omega_c(r_\beta)\cos(kz).
\label{corP02}
\end{equation}
Resonances between radial oscillations and longitudinal motion are
absent in the hamiltonian (\ref{corP02}) so the particles move
adiabatically. Condition of applicability of this approximation
$|r-r_\beta|\ll r_\beta$ can be written in the form
\begin{equation*}
\frac{2\varepsilon}{m}-\frac{p_z^2}{m^2}-\frac{(p_\theta-e\Psi_0(r_\beta)/c)^2}{m^2r_\beta^2}\ll\Omega_\beta^2r_\beta^2.
\label{corP03}
\end{equation*}
This condition means that frequency of radial oscillations (which is
the betatron frequency) have to be large enough.

\subsubsection{Particles with $\rho\ll a$.}

If radius of the diamagnetic ``bubble'' $a$ is much greater than
width of transition layer and ``Larmor radius'' $\rho$ that motion
of particles can be described approximately as motion of particle
inside surface rotation $r=r_b(z)$. Here function $r_b(z)$ is
coordinate of magnetic field line at boundary of the bubble.

Let's found the criterion of adiabaticity of particle moving with
velocity $v_0$ inside corrugated cylinder $r=a+\delta a\cos(kz)$
when corrugation is small, $\delta a\ll a$ and $k\delta a\ll1$.
Particle dynamic is described by twist mapping (see Appendix A)
\begin{eqnarray}
v_{\|n+1}=v_{\|n}-\frac{1}{k}\frac{\partial
G(v_{\|n+1},z_n)}{\partial z_{\|n}},\quad kz_{n+1}=kz_n+2k\Delta
r_{n+1}I_{n+1}+\frac{\partial G(v_{\|n+1},z_n)}{\partial v_{\|n+1}},\nonumber\\
G(v_{\|n+1},z_n)=-2\frac{\delta a}{a}k\Delta
r_{n+1}\frac{v_{\|n+1}}{I_{n+1}}\cos(kz_n+k\Delta r_{n+1}I_{n+1}),
\label{corS00}
\end{eqnarray}
here $v_{\|n}$ and $z_n$ are longitudinal velocity and coordinate at
times when radial component of velocity is zero,
$I_n=v_{\|n}/(v_0^2-v_{\|n}^2)^{1/2}$ is the ratio of longitudinal
and transversal components of velocity, $\Delta
r_n=(a^2-r_{\min}^2v_0^2/(v_0^2-v_{\|n}^2))^{1/2}$ is amplitude of
radial oscillations and $r_{\min}=|p_\theta|/(mv_0)$.

After linearization near resonances $v_\|=V_j$ (here $V_j$ are
solutions of equation $2k\Delta r_{n+1}I_{n+1}=2\pi j$ with integer
$j$) and transition to new variable
$J_n^{(j)}=(v_{\|n}-V_j)\{2k(a^2-(2(v_0^2-V_j^2)/V_j^2+1)r_{\min}^2)\}(v_0/V_j)^3\Delta
r_n^{(j)}/v_0$ one can write the twist mapping (\ref{corS00}) in the
form of the Chirikov standard map
\begin{eqnarray*}
J_{n+1}^{(j)}=J_n^{(j)}+K\sin(kz_n+\pi j+\pi),\quad
kz_{n+1}=kz_n+J_{n+1}^{(j)},\nonumber\\
K=4\frac{\delta
a}{a}\frac{v_0^2}{v_0^2-V_j^2}k^2a^2\left(1-\frac{r_{\min}^2}{a^2}\frac{v_0^2+V_j^2}{v_0^2-V_j^2}\right),
\label{corS01}
\end{eqnarray*}
here $K$ is so-called stochasticity parameter. To estimate magnitude
of corrugation needed for destroying adiabatic invariant we use the
Chirikov criterion of resonances overlapping
\begin{equation}
K>1 \label{corS02}
\end{equation}
and estimation of magnitude of corrugation of boundary of
diamagnetic ``bubble'' found in MHD approximation \cite{Khristo19}
\begin{equation}
\frac{\delta a}{a}=\frac{\delta
b}{ka}I_0(ka)\left(\frac{I_1(ka)}{I_0(ka)}+\frac{K_1(ka)}{K_0(ka)}\right).
\label{corS03}
\end{equation}
For small-scale perturbations with $ka>2$ estimation (\ref{corS03})
can be simplified:
\begin{equation}
\frac{\delta a}{a}\approx\frac{2\delta b}{ka}I_0(ka). \label{corS04}
\end{equation}

We combine expressions (\ref{corS02}) and (\ref{corS04}) to found
criterion of adiabaticity of motion in the diamagnetic trap for
particles with $\rho\ll a$:
\begin{equation}
8\delta
bkaI_0(ka)\frac{v_0^2}{v_\perp^2}\left(1-\frac{r_{\min}^2}{a^2}\frac{2v_0^2-v_\perp^2}{v_\perp^2}\right)<1,
\label{corS07}
\end{equation}
here $v_\perp=(v_0^2-v_\|^2)^{1/2}$.

This condition is first broken for particles with zero angular
momentum, $r_{\min}=0$. Most dangerous are small-scale perturbations
with $ka\gg1$, but magnitude of perturbations with very large $k$ is
seems to be small due to finite Larmor radius effects which is
neglected in expression (\ref{corS03}). Particles with great value
of $-p_\theta$ move adiabatically which consistent with results of
previous section. Particles with great value of $p_\theta$ move in
region outside the bubble where magnetic field is strong so this
particles moves adiabatically also. This behavior consistents with
criterion (\ref{corS07}).

\section{Lifetime of unconfined particles}

Let's us now looks particles which move chaotically and are not
confined absolutely. If particle moves regularly than particle
arrival to mirror at the same longitudinal velocity after each
excursion from mirror to mirror. If radial adiabatic invariant not
conserves than longitudinal velocity changes chaotically with each
approach to mirror so particle leave the diamagnetic trap after
several excursions from mirror to mirror. To estimate lifetime of
chaotically moving particles we consider population of particles
with same energy $\varepsilon$ and angular momentum $p_\theta$ which
move inside quasi-cylindrical diamagnetic ``bubble'' with radius $a$
and length $L$. Distribution function of this particles is
$f(\vv,\vr)=\delta(mv^2/2+e\varphi-\varepsilon)\delta(m(xv_y-yv_x)+e\Psi/c-p_\theta)H(v_z^2-V_{z0}(\varepsilon,p_\theta,z)^2)$,
here $V_{z0}$ is solution of equation
$I_r(\varepsilon,p_\theta,V_{z0},z)=I_r(\varepsilon,p_\theta,v_{z0},z=0)$,
$v_{z0}$ is value of longitudinal velocity corresponding margin of
adiabaticity. Full number of particles in the trap $N$ and particle
flow through mirrors $J$ are approximately
\begin{equation*}
N=\pi L\int_0^\infty dr^2\int d^3vf(\vv,r,z),\quad J=2\pi
\int_0^\infty dr^2\int_0^\infty dv_z\int dv_rdv_\varphi
v_zf(\vv,r,z_m), \label{tau00}
\end{equation*}
here $z_m$ is coordinate of the mirror. Particle lifetime is
$\tau=N/J$.

In the simplest case when all particles with energy $\varepsilon$
and angular momentum $p_\theta$ move chaotically, $V_{z0}=0$,
``bubble'' radius is large, $a\gg\rho$, and electrostatic potential
is neglectingly small, estimation of particle confinement time is:
\begin{eqnarray}
\tau\sim R_v\tau_b\left\{\begin{array}{cc}\left(a-r_{\min}\right)/\rho, & \Omega p_\theta\geq0, \\
\left(a-r_{\min}\right)/\left(\rho-2R_vr_{\min}\right), & \Omega
p_\theta<0,
\end{array}\right. \label{tau02}
\end{eqnarray}
here $\tau_b=L/(\Omega\rho)$ is of the order of particle transit
time from mirror to mirror (period of bounce-oscillations).

\section{Estimation of plasma lifetime in gas-dynamic regime}

Calculation of particle confinement time in regime of diamagnetic
confinement requires sophisticated calculations including solving of
kinetic equation together with equilibrium equation. This
calculation can be essentially simplified if plasma is quite dense
and angular scattering time is less than lifetime of unconfined
particles. In this case particle distribution functions are locally
Maxwellian and one can estimate particle confinement time by
calculating full number of particles and their flow through mirrors.

Let's assume that distribution function of particles of type $s$
approximately coincides with distribution of Maxwellian particles
inside cylinder with radius $a$
\begin{equation}
f_i(\varepsilon,p_\theta)=n_{i0}\left(\frac{m}{2\pi
T_i}\right)^{3/2}e^{-\varepsilon/T_i}H(2m_ia^2\varepsilon-p_\theta^2).
\label{GDT00}
\end{equation}
This distribution allows particle density to be uniform inside the
diamagnetic ``bubble'' with radius $a$, like in MHD models (see next
section). One can estimate lifetime of particles with distribution
$s$ as
\begin{equation}
\tau_s\sim\frac{R_vL}{v_s}\frac{a}{\rho_s}, \label{GDT01}
\end{equation}
here $v_s=(2T_s/m_s)^{1/2}$ is thermal velocity of particles of type
$s$, $\rho_s=v_s/\Omega_s$ is mean Larmor radius calculated by
vacuum magnetic field. Estimation (\ref{GDT01}) gives same lifetime
for ions and electrons with same temperatures so plasma outflow in
such regime should not be accompanied by appearing essential
ambipolar potential.

Combination $R_vL/v_i$ is of the order of time of gas-dynamic
outflow from trap with vacuum magnetic field $\tau_{GDT}$. So
transition to regime of diamagnetic confinement increases $a/\rho_i$
times particle confinement time in gas-dynamic regime. This
estimation should be compared with estimation
\begin{equation*}
\tau=\tau_{GDT}\frac{a}{\lambda} \label{GDT02}
\end{equation*}
of particle confinement time in MHD model
\cite{Beklemishev16,Khristo19}, here $\lambda$ is thickness of
boundary layer in MHD approximation. This comparison demonstrates
that kinetic effects are important when $\rho_i>\lambda$.

It should be noted that effects of adiabatic confinement of part of
particles not taken into account in estimation (\ref{GDT01}). This
effects can be important if plasma flow through mirror is
collisionless (``short'' mirrors). Effects of adiabatitity of motion
are seems to decrease particle losses. In this sense the estimation
(\ref{GDT01}) is most pessimistic estimation of particle confinement
time in the gas-dynamic regime.

\section{Numerical example}

To illustrate influence of structure of magnetic system on
collision-less particle dynamic some results of numerical simulation
of ions movement in diamagnetic trap are presented in this section.
Magnetic flux is calculated similarly article \cite{Rostoker02}.
Namely, magnetic flux satisfies Amperes's law
$(\nabla\times\vB)_\theta=\partial_rr^{-1}\partial_r\Psi=4\pi
j_\theta/c$ (here $j_\theta$ is azimuthal component of plasma
current) which can be write in the form of integral equation
\begin{eqnarray}
\Psi(r,z)=\Psi_v(r,z)+\frac{4\pi}{c}\int
dr_0dz_0\psi_G(r,r_0;z,z_0)j_\theta(r_0,z_0),\nonumber\\
j_\theta(r_0,z_0)=\sum_{s=i,e}e_s\int\frac{p_\theta-e_s\Psi(r_0,z_0)/c}{r_0}f_s\frac{dp_rdp_\theta
dp_z}{r_0}, \label{num00}
\end{eqnarray}
here $\Psi_v(r,z)$ is flux of vacuum magnetic field and
\begin{equation*}
\psi_G(r,r_0;z,z_0)=\sqrt{(z-z_0)^2+(r+r_0)^2}\frac{E(\xi)+(\xi/2-1)K(\xi)}{2\pi},
\label{num01}
\end{equation*}
is Green function (magnetic flux of thin coil), $E(\xi)$ and
$K(\xi)$ are complete elliptic integrals of first and second kinds,
$\xi=4rr_0/((z-z_0)^2+(r+r_0)^2)$.

\begin{figure}[!h]
\begin{center}
\includegraphics[width=0.45\textwidth]{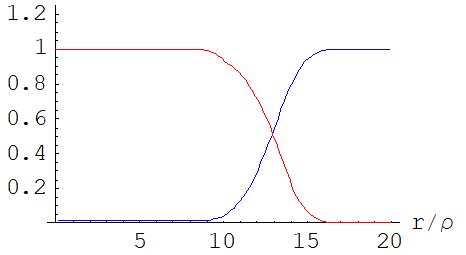}
\end{center}
\caption{\label{numN} An example of dependence of plasma density
$n/n_{i0}$ (red) and magnetic field $B_z/B_0$ (blue) on radius in
trap center. Parameters: $(2T_i/(m_i\Omega^2))^{1/2}=2$, $a=20$,
vacuum mirror ratio $R_v=2$.}
\end{figure}

The integral equation (\ref{num00}) can be solved numerically by
iterations. To calculate plasma current $j_\theta$ we assume
electrons to be cold and choose distribution function of ions
(\ref{GDT00}). Dependence of density and azimuthal current of ions
on radial coordinate and magnetic flux are given in Appendix B.
Example of radial dependence of plasma density and longitudinal
component of magnetic field on radius is shown on figure \ref{numN}.
Plasma density is constant inside the ``bubble'' like in the MHD
model \cite{Beklemishev16,Khristo19}.

\begin{figure}[!h]
\begin{center}
\includegraphics[width=0.45\textwidth]{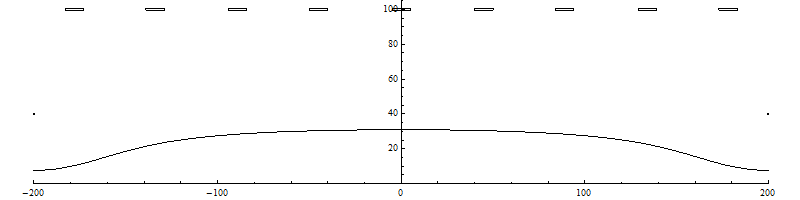}
\includegraphics[width=0.45\textwidth]{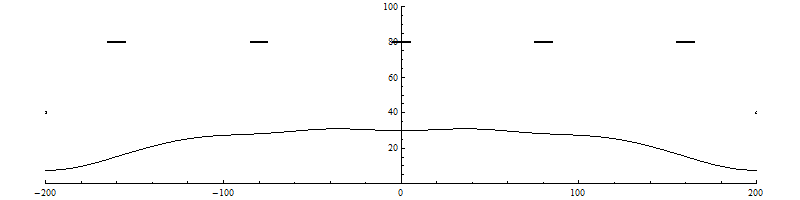}
\end{center}
\caption{\label{numB} An examples of magnet coils (rectangles) and
magnetic field line on ``bubble'' boundary (solid curve) for smooth
(left) and corrugated (right) vacuum magnetic field. Parameters:
$(2T_i/(m_i\Omega^2))^{1/2}=2$, $a=20$.}
\end{figure}

Magnetic system of trap consists of two mirror coils and of set of
equidistant coaxial coils which generate quasi-uniform magnetic
field (see figure \ref{numB}). In one case corrugation of vacuum
magnetic field does not exceeds tenths of percent (smooth magnetic
field). In second case distance between the coils is doubled and
radius of the coils is reduced (currents in coils are changed
correspondingly so that value of magnetic field in center is the
same in both cases). It results in observable corrugation of the
``bubble'' boundary.

\begin{figure}[!h]
\begin{center}
\includegraphics[width=0.6\textwidth]{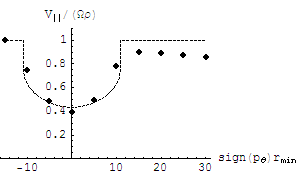}
\end{center}
\caption{\label{numVl} An example of value of critical longitudinal
velocity for ions with $\rho=2$ moving in corrugated magnetic field
(points) and margin of adiabaticity (\ref{corS07}) at $ka=2.7$ and
$\delta b=0.01$ (dashed line).}
\end{figure}

Numerical simulation allows us to found maximal value of
longitudinal velocity at which ions confine in the trap. Ions move
regularly in trap with smooth field and maximal critical velocity is
restricted only by criterion of confinement (\ref{Ir03}). An example
of dependence of critical velocity on angular momentum for ions
moving in trap with corrugated field is shown on figure \ref{numVl}.
Ions with small $|p_\theta|$ scatter due to corrugation so their
critical velocity is relatively low. This velocity increases when
$|p_\theta|$ rises (in according with criterion of adiabaticity
(\ref{corS07})). Ions with negative and small $p_\theta$ are
confined absolutely so their maximal velocity is restricted by full
energy $(2\varepsilon/m_i)^{1/2}$. Critical velocity of ions with
large $p_\theta$ decreases because this ions move in region with
finite magnetic field outside the ``bubble''.

\begin{figure}[!h]
\begin{center}
\includegraphics[width=0.45\textwidth]{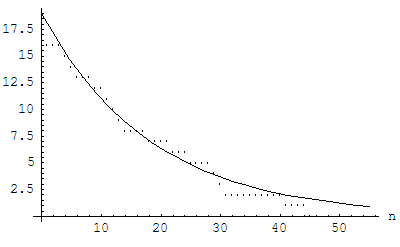}
\end{center}
\caption{\label{Time} An example of number of unconfined ions after
$n$ bounce oscillations (points) and function $19e^{-x/18.3}$
(solid). Parameters: $\rho=2$, $r_{\min}=0$.}
\end{figure}

An example of number of unconfined ions in trap with corrugated
field after $n$ bounce-oscillations is shown on figure \ref{Time}.
This number decreases approximately exponentially with time.
Confinement time $18.3\tau_b$ is consistent with analytical
estimation (\ref{tau02}).

\section{Conclusion}

Smallness of magnetic field in central region of the diamagnetic
trap results in non-conservation of magnetic moment so regimes of
particle confinement are modified. Particle can confine inside the
diamagnetic ``bubble'' either due to conservation of radial
adiabatic invariant (this mechanism occurs if vacuum magnetic field
is smooth) or in regime of adiabatic confinement (if particle
rotates quickly around axis of trap in direction coinciding with
direction of cyclotron rotation). Possibility of conservation of the
adiabatic invariant depends strongly on geometry of magnetic field,
especially on small-scale perturbation of vacuum magnetic field.
Lifetime of unconfined particles increases with increasing the
``bubble'' radius and vacuum magnetic field in the mirrors of the
trap and decreasing particle ``Larmor radius'' $\rho$. Even in the
worst case when all particles move chaotically particle confinement
time exceeds the gas-dynamic time in the vacuum field in ratio of
the ``bubble'' radius to mean ion ``Larmor radius''.

The author wish to thank all the members of the laboratories 9-0,
9-1 and 10 of BINP SB RAS who participated in discussion of the
results of this work. Author especially grateful to the Dr. Alexei
Beklemishev, Dr. Dmitriy Skovorodin and Mikhail Khristo for fruitful
discussions.

\section{Appendix A. Twist mapping for particle inside corrugated surface}

Now we consider particle moving with velocity $v_0$ inside
corrugated cylindrical surface and reflecting elastically from it.
Let $z_n$ and $v_{\|n}$ to be longitudinal coordinate and
longitudinal component of velocity of particle at time $t_n$. At
this point of time radial velocity equal to zero and azimuthal
component of velocity is $(v_0^2-v_{\|n}^2)^{1/2}$. Radial
coordinate of particle is $r_n=r_{\min}v_0/(v_0^2-v_{\|n}^2)^{1/2}$,
here $r_{\min}=|p_\theta|/(mv_0)$. Particle will collide with
surface at point of time $t_w$ which is solution of equation
$(r_n^2+(v_0^2-v_{\|n}^2)t_w^2)^{1/2}=a+\delta
a\cos(kz_n+kv_{\|n}t_w)$. Longitudinal and radial coordinate of the
particle at the moment of collision are $z_w=z_n+v_{\|n}t_w$ and
$r_w=a+\delta a\cos(kz_w)$. Radial, azimuthal and longitudinal
components of velocity at the moment of collision are
$v_r=(v_0^2-v_0^2r_{\min}^2/r_w^2-v_{\|n}^2)^{1/2}$,
$v_\theta=v_0r_{\min}/r_w$ è $v_\|=v_{\|n}$. After collision radial
and longitudinal components of velocity are
$v_{\|n+1}=(2f'v_r+(1-f'^2)v_{\|n})/(1+f'^2)$ è
$v_r'=(2f'v_{\|n}-(1-f'^2)v_r)/(1+f'^2)$, here $f'=-k\delta
a\sin(kz_w)$. Azimuthal component does not change. Radial coordinate
of particle will reach minimal value
$r_{n+1}=r_{\min}v_0/(v_0^2-v_{\|n+1}^2)^{1/2}$ through time $\Delta
t=(r_w^2-r_{\min}^2v_0^2/(v_0^2-v_{\|n+1}^2))^{1/2}/(v_0^2-v_{\|n+1}^2)^{1/2}$
after collision. Longitudinal velocity is
$z_{n+1}=z_n+v_{\|n}t_w+v_{\|n+1}\Delta t$ when radial coordinate is
minimal.

If corrugation is weak $\delta a\ll a$ that time before collision is
approximately $t_w\approx(\Delta r_nI_n/v_{\|n})\{1+(\delta
aa/\Delta r_n^2)\cos(kz_n+k\Delta r_nI_n)\}$, here $\Delta
r_n=(a^2-r_{\min}^2v_0^2/(v_0^2-v_{\|n}^2))^{1/2}$ and
$I_n=v_{\|n}/(v_0^2-v_{\|n}^2)^{1/2}$. When radial coordinate
minimal longitudinal component of velocity and longitudinal
coordinate of particle are described by expressions (\ref{corS00}).

\section{Appendix B. Density and current of ions.}

Density of ions with distribution function (\ref{GDT00}) is
\begin{eqnarray*}
n_i/n_{i0}=H(r_b-r)e^{-2e\varphi/(mw^2)}+H(r-r_b)\frac{r_b}{r}+\nonumber\\
+\frac{H\left(\Omega^2\psi^2-2(r_b^2-r^2)e\varphi/m_i\right)}{2}\left(\frac{r_b}{r}\left\{\erf(\frac{p_+/m_i}{r_bw})-\erf(\frac{p_-/m_i}{r_bw})\right\}-\right.\nonumber\\
\left.-e^{-2e\varphi/(mw^2)}\left\{\erf(\frac{p_+/m_i-\Omega\psi}{rw})-\erf(\frac{p_-/m_i-\Omega\psi}{rw})\right\}\right),
\label{ApB00}
\end{eqnarray*}
here $\psi(r,z)=\Psi(r,z)/B_0$ is normalized magnetic flux,
$\varphi$ is electrostatic potential, $w=(2T_i/m_i)^{1/2}$ is
thermal velocity and
\begin{equation*}
p_\pm=m_i\frac{\Omega r_b^2\psi\pm
r_br\sqrt{2(r^2-r_b^2)e\varphi/m_i+\Omega^2\psi^2}}{r_b^2-r^2}
\label{ApB01}
\end{equation*}

Current of ions is
\begin{eqnarray*}
j_i/(en_{i0})=-H(r-r_b)\Omega\psi\frac{r_b}{r^2}+\frac{H\left(\Omega^2\psi^2-2(r_b^2-r^2)e\varphi/m_i\right)}{2}\times\nonumber\\
\times\left(e^{-2e\varphi/(mw^2)}\frac{w}{\sqrt\pi}(e^{-(p_+/m_i-\Omega\psi)^2/(r^2w^2)}-e^{-(p_-/m_i-\Omega\psi)^2/(r^2w^2)})+\right.\nonumber\\
\left.+\frac{r_b}{r^2}\left\{(e^{-(p_+/m_i)^2/(r_b^2w^2)}-e^{-(p_-/m_i)^2/(r_b^2w^2)})\frac{wr_b}{\sqrt\pi}+\Omega\psi(\erf(\frac{p_+/m_i}{r_bw})-\erf(\frac{p_-/m_i}{r_bw}))\right\}\right).
\label{ApB02}
\end{eqnarray*}

\par

\end{document}